\documentclass[aps,
preprint,superscriptaddress,groupedaddress]{revtex4}

\usepackage{graphicx}
\usepackage{amsmath}
\usepackage{latexsym}      
\usepackage{anysize} 
\usepackage{array}
\usepackage[T1]{fontenc} 

\begin{document}

\title{Epidemic spreading and immunization strategy in multiplex
  networks}

\author{Lucila G. Alvarez Zuzek} \email{lgalvere@mdp.edu.ar}
\affiliation{Departamento de F\'{i}sica, Facultad de Ciencias Exactas
  y Naturales, Universidad Nacional de Mar del Plata, and Instituto de
  Investigaciones F\'{\i}sicas de Mar del Plata (IFIMAR-CONICET),
  De\'an Funes 3350, 7600 Mar del Plata, Argentina} 

\author{Camila Buono} \affiliation{Departamento de F\'{i}sica,
  Facultad de Ciencias Exactas y Naturales, Universidad Nacional de
  Mar del Plata, and Instituto de Investigaciones F\'{\i}sicas de Mar
  del Plata (IFIMAR-CONICET), De\'an Funes 3350, 7600 Mar del Plata,
  Argentina} 

\author{Lidia A. Braunstein} \affiliation{Departamento de F\'{i}sica,
  Facultad de Ciencias Exactas y Naturales, Universidad Nacional de
  Mar del Plata, and Instituto de Investigaciones F\'{\i}sicas de Mar
  del Plata (IFIMAR-CONICET), De\'an Funes 3350, 7600 Mar del Plata,
  Argentina}\affiliation{Center for Polymer Studies, Boston
  University, Boston, Massachusetts 02215, USA.}

\begin{abstract}
A more connected world has brought major consequences such as
facilitate the spread of diseases all over the world to quickly become
epidemics, reason why researchers are concentrated in modeling the
propagation of epidemics and outbreaks in multilayer networks. In this
networks all nodes interact in different layers with different type of
links. However, in many scenarios such as in the society, a multiplex
network framework is not completely suitable since not all individuals
participate in all layers. In this paper, we use a {\it partially
  overlapped multiplex} network where only a fraction of the
individuals are shared by the layers. We develop a mitigation strategy
for stopping a disease propagation, considering the
Susceptible-Infected-Recover model, in a system consisted by two
layers. We consider a random immunization in one of the layers and
study the effect of the overlapping fraction in both, the propagation
of the disease and the immunization strategy. Using branching theory,
we study this scenario theoretically and via simulations and find a
lower epidemic threshold than in the case without strategy.
\end{abstract}

\maketitle

\section{Introduction}

In the last years the complex networks analysis has been focused in no
further considering networks as isolated entities, but characterizing
how networks interact with other networks and how this interaction
affects processes that occurs on top of them. A system composed of
interconnected networks is called a \emph{Network of Networks\/} (NoN)
\cite{jia_02,Gao_12,Gao_01,Val13}. In NoN there are connectivity links
within each individual network, and external links that connect each
network to other networks in the system. Very recently physicists have
begun to consider a particular class of NoN in which the nodes have
multiple types of links across different \emph{layers\/} called {\it
  multiplex or multilayer} networks
\cite{Lee_12,Brummitt_12,Gomez_13,Kim_13,Cozzo_12,GoReArFl12,Kiv_13}.

Recently, the study of the effect of multiplexity of networks in
propagation processes such as epidemics has been the focus of many
recent researches \cite{Dickison_12,Mar_11,Yag_13,Cozzo_13}. In
Ref~\cite{Buo_14} the research concentrated in the propagation of a
disease in partially overlapped multilayer networks, owing to the fact
that individuals are not necessarily present in all the layers of a
society and this has an impact in the epidemic propagation. For the
epidemic model they used the susceptible-infected-recovered (SIR)
model \cite{Bailey_75,Colizza_06,Colizza_07} that describes the
propagation of non recurrent diseases in which infected individuals
either die or, after recovery, become immune to future infections. In
the SIR model each individual of the population can be in one of three
different states: Susceptible, Infected, or Recovered. Infected
individuals transmit the disease to its susceptible neighbors with a
probability $\beta$ and recover after a fixed period of time
$t_r$. The spreading process stops when there is only susceptible
and/or recovered nodes. The dynamic of the epidemic is controlled by
the transmissibility $T=1-(1-\beta)^{t_r}$, which is a measure of the
disease virulence, i.e., the effective probability that the disease
will be transmitted by an infected individual across any given
link. At the final state of this process, the fraction of recovered
individuals $R$ is the order parameter of a second order phase
transition with a control parameter $T$. For $T < T_c$, where $T_c$ is
the epidemic threshold, there is an epidemic-free phase with only
small outbreaks. However, for $T \geq T_c$, an epidemic phase
develops. In isolated networks the epidemic threshold is given by $T_c
= 1/(\kappa-1)$, where $\kappa$ is the branching factor that is a
measure of the heterogeneity of the network.  The branching factor is
defined as $\kappa = \langle k^2 \rangle / \langle k \rangle$, where
$\langle k^2 \rangle$ and $\langle k \rangle$ are the second and first
moment of the degree distribution, respectively. Since the SIR model
presents a local tree structure we can employ the branching theory
approach within a generating function formalism \cite{Dun_01,New_03}
that holds in the thermodynamic limit.  In \cite{Buo_14} the SIR model
was studied, with $\beta$ and $t_r$ constant, in a system composed of
two overlapping layers in which only a fraction $q$ of individuals can
act in both layers. In their model, the two layers represent contact
networks in which only the overlapping nodes enable the propagation
between layers, and thus the transmissibility $T$ is the same in both
layers.  They found that decreasing the overlap decreases the risk of
an epidemic compared to the case of full overlap ($q=1$). They also
found that the critical threshold increases as $q$ decreases, and that
in the limit of small overlapping fraction, the epidemic threshold is
dominated by the most heterogeneous layer, this effect could have
important implications in the implementation of mitigation strategies.

Motivated by this, in this work we study a disease spreading process
in overlapped multiplex networks and an immunization strategy for the
epidemic spreading. For the strategy, we use a random immunization of
individuals in one layer of the network. Those immunized overlapped
individuals will remain immunized in all layers of the network.

\section{Epidemic propagation process}

\begin{figure}[h]
  \begin{center}
 \includegraphics[scale=0.75]{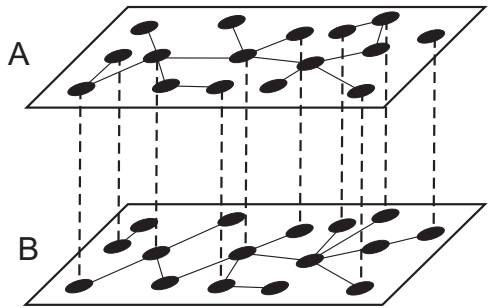}
  \end{center}
  \caption{Partially overlapped multiplex
    network with layer size $N=16$ and fraction of shared nodes
    $q=0.625$. The total size of the network is $(2-q)N=22$
    individuals. The dashed lines are used as a guide to show the
    fraction $q$ of shared nodes. Before the spreading dynamics, all
    individuals are in the susceptible stage represented by black
    circles.}
  \label{NetGM}
\end{figure}

In our model we use an overlapping multiplex network formed by two
layers, $A$ and $B$, of the same size $N$, where an overlapping fraction
$q$ of {\it shared\/} individuals is active in both layers.
Figure~\ref{NetGM}(a) shows schematically the partially overlapped
network. The dashed lines that represent the fraction $q$ of shared
individuals should not to be interpreted as interacting or
interdependent links but as the shared nodes and their counterpart in
the other layer.

For the simulation, we construct each layer using the Molloy Reed
algorithm \cite{Mol_95}, and we choose randomly a fraction $q$ of
nodes in each of the layers that represent the same nodes. In our
model we assume that the transmissibility is the same in both layers
because there is only one disease and all individuals in the system
spread with the same probability. We begin by infecting a randomly
chosen individual in layer $A$. The spreading process then follows the
SIR dynamics in both layers, the overlapped nodes in both layers have
the same state because they represent the same individuals. After all
infected nodes infect their susceptible neighbors with probability
$\beta$ in both layers, the time is increased in one, and the states
of the nodes are updated simultaneously. Note that because there are
shared nodes the branches of infection can cross between the two
layers.  Thus the probability that, following a random link, a node
belonging to the infected branch will be reached in each layer can be
written as,

\begin{eqnarray}
  f_A  = (1-q)\; [1-G_1^A(1-Tf_A)]+q \;[1-G_1^A(1-Tf_A)\;G_{0}^B(1-Tf_B)] \; , \label{fA} \\
  f_B  = (1-q)\; [1-G_1^B(1-Tf_B)]+q \;[1-G_1^B(1-Tf_B)\;G_{0}^A(1-Tf_A)]  \; ,
  \label{fB}
\end{eqnarray}
where $G_0^i(x)=\sum_{k=k_{\rm min}}^{k_{\rm max}} P_i(k) x^{k}$ is the
generating function of the degree distribution and
$G_1^i=\sum_{k=k_{\rm min}}^{k_{\rm max}} P_i (k) \; k
x^{k-1}$ is the generating function of the excess degree
distribution in layer $i$, with $i=A,B$ \cite{New_03}.

Equation (\ref{fA}) has two terms, since the probability $f_A$ to
expand an infected branch following a random chosen link in layer $A$,
can be written as the probability to reach one of the $(1-q)$
non-overlapped individuals and that the branch of infection expands
through the $k-1$ remaining connections of the individual in layer $A$,
combined with the probability of reaching one of the $q$ overlapped
individuals and that the branch of infection expands through the $k-1$
remaining connections of the individual in layer $A$ and through the $k$
connections of the individual in layer $B$. An analogous interpretation
holds for the equation (\ref{fB}).

The solution of the system of equations (\ref{fA}) and (\ref{fB}) for
all $T$ above and at criticality is given by the intersection of the
curves $f_A$ and $f_B$. At criticality, this intersection can be
derived by solving the determinant equation $|J-I|=0$, where $I$ is
the identity and $J$ is the Jacobian matrix of the system of equations
(\ref{fA}) and (\ref{fB}). The only possibility to have a non-epidemic
regime is that none of the branches of infection spread, {\it i.e.}
$f_A=f_B=0$, therefore below and at criticality $f_A=f_B=0$. The
evaluation of the Jacobian matrix $J_{ij} = (\partial f_{i} / \partial
f_{j})|_{f_A=f_B=0}$ allow us to obtain a quadratic equation for $T_c$
with only one stable solution \cite{All_97} given by,

\begin{equation}
T_c = \frac{[(\kappa_A-1)+(\kappa_B-1)] - \sqrt{[(\kappa_A-1)-(\kappa_B-1)]^2
    + 4 q^2 \langle k_A\rangle \langle k_B\rangle}}{2
  (\kappa_A-1)(\kappa_B-1) - 2q^2 \langle k_A\rangle \langle k_B\rangle},
\label{tc}
\end{equation}
where $\kappa=1+1/T_c$ is the total branching factor of the system and
$\kappa_A$, $\kappa_B$ are the isolated branching factors of layer $A$
and $B$ respectively. For $q \to 0$ we recover the isolated network
result $T_c=1/(\kappa_A -1)$, which is compatible with our model in
which the infection starts in layer $A$ and  the disease never reaches
layer $B$. In contrast, when $q\to 1$, we find that $T_c =
1/\sqrt{[(\kappa_A-\kappa_B)]^2+4\langle k_A\rangle \langle
  k_B\rangle}$. Note that $T_c(q\to 1) < T_c(q\to0)$. In general,
$T_c$ decreases with $q$. This is the case because an increase in the
overlapping between layers increases the total branching factor, and
therefore the total system becomes more heterogeneous in degree, {\it
  i.e.}, the total branching factor is equal to or bigger than the
branching factor of the isolated layers.

\section{Immunization strategy}

We study a random immunization strategy on the partially overlapped
multiplex network. We start by immunizing a random fraction $m$ of
individuals in layer $A$, before the epidemic spreading take place. An
immunized individual will be immune to the disease in all layers, and
therefore can not be infected or infect during all the propagation
process. Note that, due to the presence of the overlapped individuals,
in layer $B$ there will be a random fraction $mq$ of immunized
individuals.

After immunizing, we spread a disease in the network, starting by
infecting a random susceptible non-immunized individual in layer $A$
(patient zero). Thus, the probability that reaching a node by
following a randomly chosen link, it belongs to a branch of infection
is given by the system of equations (\ref{fA}) and (\ref{fB}), using a
node diluted degree distribution in each layer \cite{Coh_10} due to
the immunization strategy. Thus with the diluted degree distribution
we have that the branching factor of the diluted layers are,

\begin{eqnarray}
\widetilde{\kappa_A}  & = & (1-m) \; \kappa_A \label{kappaA}\\ 
\widetilde{\kappa_B}  & = & (1-qm) \; \kappa_B\;,
\label{kappaB}
\end{eqnarray}
where $\kappa_A$ and $\kappa_B$ are the branching factor of the
original layers respectively. Note that the branching factor is
reduced due to the immunization strategy increasing the epidemic
threshold and thus hindering the diseases propagation.
\\
\\
\begin{figure}[h]
  \begin{center}
  \includegraphics[scale=0.4]{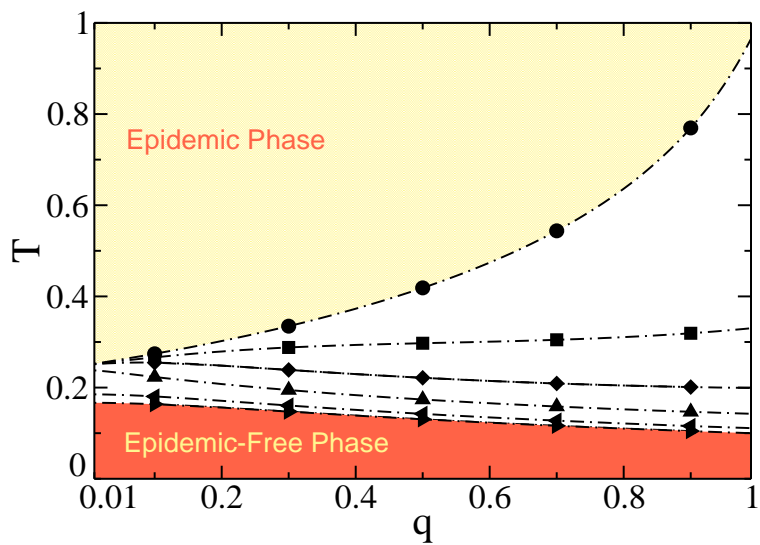}
  \end{center}
  \caption{Phase diagram in the plane $T-q$ for the SIR model in the
    multiplex network, when the random immunization strategy is
    applied, for different values of the immunized fraction $m$. Both,
    layer $A$ and $B$, have Erd\H{o}s R\'enyi degree distributions
    with mean values of connectivity $\langle k_A \rangle=6$ and
    $\langle k_B \rangle=4$ for layer $A$ and $B$ respectively. Symbols
    corresponds to the value of $T_c$ for different values of $m$
    obtained by numerical simulation with layer size $N=10^5$, while
    the lines denote the theoretical results obtained numerically from
    Eqs. (\ref{fA}) and (\ref{fB}) using $\widetilde{\kappa_A}$ and
    $\widetilde{\kappa_B}$ given by Eqs. (\ref{kappaA}) and
    (\ref{kappaB}). From top to bottom $m=0.9;0.7;0.5;0.3;0.1;0$. Above
    the lines the system is in the epidemic phase for each value of
    $m$, and below it is in the epidemic-free phase where the disease
    can not propagate. All simulations were done over $10^5$ network
    realizations.}
  \label{Tvsq}
\end{figure}

In Figure~\ref{Tvsq} we show the phase diagram in the plane $T-q$ for
different values of the immunization fraction $m$. We consider that
both layers have Erd\H{o}s R\'enyi degree distributions with mean
values of connectivity $\langle k_A \rangle=6$ and $\langle k_B
\rangle=4$ for layer $A$ and $B$ respectively, and we use $k_{min} = 1$
and $k_{max}=40$ as the minimum and maximum connectivity respectively
in each layer. The lines represent $T_c$ for many values of $m$
obtained theoretically from Eqs. (\ref{fA}) and (\ref{fB}) while
symbols denote the numerical simulation results. Above $T_c$ there is
an epidemic phase and below $T_c$ only outbreaks exists (non-epidemic
phase). Fig. \ref{Tvsq} shows that $T_c$ has different behaviors with
$q$ depending on the value of $m$. From Figure~\ref{Tvsq} we can see a
good agreement between the theoretical predictions and the numerical
simulation results.

Note that when $q=0$ (not shown) the critical threshold corresponds to
an isolated layer in which the disease starts, {\it i.e.} layer $A$
and where the critical threshold is given by $T_c = 1/(\kappa_A
-1)$. For $q \to 0$ ($q=0.01$) the epidemic threshold converges to the
threshold of the layer with the bigger branching factor, since in this
limit the process is dominated by the most heterogeneous layer
\cite{Buo_14}. We can observe from Fig. \ref{Tvsq} that as the
immunization fraction increases, the epidemic-free phase widens. When
$m < 0.7$ (see Fig. \ref{Tvsq}) $T_c$ decreases with $q$ owing to the
fact that as the overlapping between the layers increases the total
branching factor of the network increases. However, for $m \geq 0.7$
$T_c$ increases as $q$ increases. This last effect can be understood
taking into account that for $m > 0.7$ layer $A$ is very diluted, thus
the disease spreads mostly through layer $B$, as $q$ increases the
immunization fraction $mq$ of layer $B$ increases, hindering the
propagation through that layer. It is expected that for more
heterogeneous networks this strategy has less impact in the spreading
process, due to the fact that the more heterogeneous the network is,
the more harder it is to dilute with this strategy.

\section{Discussion}

In this work we study, theoretically and via simulations, an epidemic
spreading and a random immunization strategy in a partially overlapped
multiplex network composed by two layer with an overlapping fraction
$q$. We immunize a fraction $m$ of individuals in one layer of the
network and study how this process affects the propagation of the
disease through all layers. We found that for $q \to 0$ the critical
threshold of the epidemic is dominated by the threshold of the most
heterogeneous layer for all $m > 0$. We found that there is a regime
in which $T_c$ decreases with $q$ due to the fact that the total
branching factor of the system increases. This behavior stands for $m
< 0.7$, however for bigger values of $m$, $T_c$ increases as $q$
increases, hindering the disease propagation. This last effect can be
understood taking into account that when $m>0.7$, layer $A$ is diluted,
and as $q$ increases the immunization fraction $mq$ of layer $B$
increases, and the effect of the immunized individuals in that layer
is stronger.

We can observe from Fig. \ref{Tvsq} that as the immunization fraction
increases, the epidemic-free phase widens. When $m < 0.7$ we can see
that $T_c$ decreases with $q$ owing to the fact that as the
overlapping between the layers increases the total branching factor of
the network increases. However, for $m \geq 0.7$, $T_c$ increases as
$q$ increases, hindering the disease propagation. This last effect can
be understood taking into account that as $q$ increases the
immunization fraction $mq$ of layer $B$ increases and for $m>0.7$ the
effect of the immunized individuals in that layer is stronger.  Our
study suggests that vaccinating or isolating only in one layer with
the higher propagation capacity, can reduce drastically the total
branching factor of the network. As a consequence, the epidemic
threshold of the system increases significantly, reducing the risk of
a disease epidemic in the system.

\section{Acknowledgments}

The authors wish to thank to UNMdP and FONCyT (Pict 0429/2013) for
financial support. LGAZ wishes to thank CCP 2014 Sponsors for
financial support to assist to the conference. LGAZ also wishes to
thank Professor H. E. Stanley to host her in the Center for Polymer
Studies. The authors wish to thank Lucas D. Valdez for his useful
comments and discussions.

\bibliography{LGAZ}

\providecommand{\newblock}{}
\begin{thebibliography}{10}
\expandafter\ifx\csname url\endcsname\relax
  \def\url#1{{\tt #1}}\fi
\expandafter\ifx\csname urlprefix\endcsname\relax\def\urlprefix{URL }\fi
\providecommand{\eprint}[2][]{\url{#2}}

\bibitem{jia_02}
Gao J, Buldyrev S~V, Havlin S and Stanley H~E 2011 {\em Phys. Rev. Lett.\/}
  {\bf 107} 195701

\bibitem{Gao_12}
Gao J, Buldyrev S~V, Stanley H~E and Havlin S 2012 {\em Nature Physics\/} {\bf
  8}

\bibitem{Gao_01}
Dong G, Gao J, Du R, Tian L, Stanley H~E and Havlin S 2013 {\em Phys. Rev. E\/}
  {\bf 87} 052804

\bibitem{Val13}
Valdez L~D, Macri P~A, Stanley H~E and Braunstein L~A 2013 {\em Phys. Rev. E\/}
  {\bf 88} 050803(R)

\bibitem{Lee_12}
Lee K~M, Kim J~Y, Cho W~K, Goh K~I and Kim I~M 2012 {\em New J. Phys.\/} {\bf
  14} 033027

\bibitem{Brummitt_12}
Brummitt C~D, Lee K~M and Goh K~I 2012 {\em Phys. Rev. E\/} {\bf 85} 045102(R)

\bibitem{Gomez_13}
G{\'o}mez S, D{\'i}az-Guilera A, G{\'o}mez-Garde{\~n}es J, P{\'e}rez-Vicente
  C~J, Moreno Y and Arenas A 2013 {\em Phys. Rev. Lett.\/} {\bf 110} 028701

\bibitem{Kim_13}
Kim J~Y and Goh K~I 2013 {\em Phys. Rev. Lett.\/} {\bf 111} 058702

\bibitem{Cozzo_12}
Cozzo E, Arenas A and Moreno Y 2012 {\em Phys. Rev. E\/} {\bf 86} 036115

\bibitem{GoReArFl12}
G{\'o}mez-Garde{\~n}es J, Reinares I, Arenas A and Floria L~M 2012 {\em Nature
  Scientific Reports\/} {\bf 10.1038} srep00620

\bibitem{Kiv_13}
Kivel{\"a} M, Arenas A, Barthelemy M, Gleeson J~P, Moreno Y and Porter M~A 2013
  {Multilayer Networks} http://arxiv.org/abs/1309.7233

\bibitem{Dickison_12}
Dickison M, Havlin S and Stanley H~E 2012 {\em Phys. Rev. E\/} {\bf 85} 066109

\bibitem{Mar_11}
Marceau V, No{\"e}l P, H{\'e}bert-Dufresne L, Allard A and Dub{\'e} L~J 2011
  {\em Phys. Rev. E\/} {\bf 84}(2) 026105

\bibitem{Yag_13}
Yagan O, Qian D, Zhang J and Cochran D 2013 {\em IEEE JSAC Special Issue on
  Network Science\/} {\bf 31} 1038

\bibitem{Cozzo_13}
Cozzo E, Ba{\~n}os R~A, Meloni S and Moreno Y 2013 {\em Phys. Rev. E\/} {\bf
  88} 050801(R)

\bibitem{Buo_14}
Buono C, Zuzek L~G~A, Macri P~A and Braunstein L~A 2014 {\em PLoS ONE\/} {\bf
  9} e9220

\bibitem{Bailey_75}
Bailey N~T~J 1975 {\em {The Mathematical Theory of Infectious Diseases}\/}
  (Griffin, London)

\bibitem{Colizza_06}
Colizza V, Barrat A, Barthlemy M and Vespignani A 2006 {\em Proc. Natl. Acad.
  Sci. USA\/} {\bf 103} 2015

\bibitem{Colizza_07}
Colizza V and Vespignani A 2007 {\em Phys.Rev.Lett.\/} {\bf 99} 148701

\bibitem{Dun_01}
Callaway D, Newman M~E~J, Strogatz S~H and Watts D~J 2000 {\em Phys. Rev.
  Lett.\/} {\bf 85} 5468

\bibitem{New_03}
Newman M~E~J, Strogatz S~H and Watts D~J 2001 {\em Phys. Rev. E\/} {\bf 64}
  026118

\bibitem{Mol_95}
Molloy M and Reed B 1995 {\em Random structures and algorithms\/} {\bf 6} 161

\bibitem{All_97}
Alligood K~T, Sauer T~D and Yorke J~A 1997 {\em {CHAOS: An Introduction to
  Dynamical Systems}\/} (Springer)

\bibitem{Coh_10}
Cohen R and Havlin S 2010 {\em {Complex Networks: Structure, Robustness and
  Function}\/} (Cambridge University Press)

\end{thebibliography}

\end{document}